\documentclass[a4paper]{article}
\usepackage{eqis}
\input{epsf}
\begin{document}

\title{
   Effects of decoherence and imperfections 
for quantum algorithms
}

\author{
  A.A. Pomeransky
  \affiliation{1},
  \and
  O.V. Zhirov
  \affiliation{2},
  \and
  D.L. Shepelyansky
  \affiliation{1}
  \email{http://www.quantware.ups-tlse.fr}
}

\address{1}{
  Laboratoire de Physique Th\'eorique, UMR 5152 du CNRS, Univ. P. Sabatier,
  31062 Toulouse Cedex 4, France
}
\address{2}{
  Budker Institute of Nuclear Physics, 630090 Novosibirsk, Russia
}

\abstract{ 
We study effects of static inter-qubit interactions 
and random errors in quantum gates on the stability of various
quantum algorithms including 
the Grover quantum search algorithm and the quantum chaos maps.
For the Grover algorithm
our numerical and analytical results show existence of regular and 
chaotic phases depending on the static imperfection
strength $\varepsilon$. 
The critical border $\varepsilon_c$ between two phases drops polynomially 
with the number of qubits $n_q$ as $\varepsilon_c \sim n_q^{-3/2}$. 
In the regular phase $(\varepsilon < \varepsilon_c)$ the algorithm remains 
robust against imperfections showing the efficiency gain $\varepsilon_c / \varepsilon$ 
for $\varepsilon > 2^{-n_q/2}$.
In the chaotic phase $(\varepsilon > \varepsilon_c)$ the algorithm
is completely destroyed. The results for the Grover algorithm are compared
with the imperfection effects for
quantum algorithms of quantum chaos maps
where the universal law for the fidelity decay is given by the 
Random Matrix Theory (RMT). We also discuss a new gyroscopic 
quantum error correction method which allows to reduce the effect of 
static imperfections. In spite of this decay GYQEC allows to obtain
a significant gain in the accuracy of quantum computations.
}

\keywords{Imperfections, quantum chaos, random matrix theory, 
quantum error correction}

\maketitle


In realistic quantum computations \cite{chuang}
the elementary gates are never perfect and
therefore it is very important to analyze the effects of imperfections and quantum errors on 
the algorithm accuracy. A usual model of quantum errors assumes that angles of unitary
rotations fluctuates randomly in time for any qubit in some small interval $\varepsilon$ near 
the exact angle values determined by the ideal algorithm. In this case a realistic quantum
computation remains close to the ideal one up to a number of performed gates  
$N_g \sim 1/\varepsilon^2$. For example, the fidelity $f$ of computation, defined as
a square of scalar product of quantum wavefunctions of ideal and perturbed algorithms, 
remains close to unity if a number of performed gates is smaller than $N_g$. This result
has been established analytically and numerically in extensive studies of various quantum 
algorithms \cite{Paz,Frahm03}.

Another source of quantum errors comes from internal imperfections 
generated by residual static
couplings between qubits and one-qubit energy level shifts which 
fluctuate from one qubit to 
another but remain static in time. These static imperfections 
may lead to appearance of many-body
quantum chaos, which modifies strongly the hardware properties 
of realistic quantum computer
\cite{Geor00}.
The effects of static imperfections on the accuracy of 
quantum computation have been investigated 
on the examples of quantum algorithms for the models of complex quantum dynamics 
(see e.g. \cite{Frahm03} and Refs. therein)
and a universal law for fidelity decay induced by static imperfections has been 
established \cite{Frahm03} for quantum algorithms simulating dynamics in the regime of 
quantum chaos. This law is based on the RMT treatment of imperfections.
At the same time it has been realized that the effects of static imperfections for 
dynamics in an integrable regime are not universal and more complicated. 
Therefore it is important
to investigate the effects of static imperfections on an example of the well known
Grover algorithm (GA) \cite{Grov97}. The results of these investigations \cite{grover1}
show that a quantum phase transition to quantum chaos
takes place for the imperfection strength $\varepsilon > 
\varepsilon_c \approx 1.7 /(n_g\sqrt{n_{tot}})$ where $n_g = O(n_{tot})$ is the total
number of quantum gates for one Grover iteration and 
$n_{tot}=n_q+1$ is the total number of qubits.
Notations and detailed explanations are given in \cite{grover1}. Here 
we give the description of the gyroscopic quantum error correction (GYQEC) method 
allowing significantly  suppress static imperfections in GA.
For the first time this method is  discussed in \cite{alber} for the quantum tent map. 
\begin{figure}
\epsfxsize=7.5cm
\epsfysize=4.8cm
\epsffile{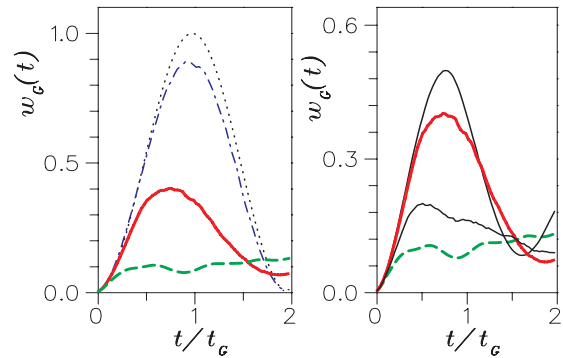}
\vglue -0.4cm
\caption{Probability $w_G(t)$ of a searched state in GA
as a function of the Grover iteration  $t$  at
$n_{tot}=12, \varepsilon= 0.002, t_G=34.5$.
Left: curves  show data for ideal GA, GA with gate to gate randomly fluctuating
coefficients $a_i, b_{ij}$ (see Eq. (2) in \cite{grover1}),
GYQEC with $l_g=10$, GA with static imperfections
(from top to bottom at $t/t_g=1$). Right: curves show data 
for GYQEC at $l_g=1,10,20$ and GA with static imperfections
(also from top to bottom). 
} 
\label{fig1}
\end{figure}

\begin{figure}
\epsfxsize=8.0cm
\epsfysize=5.2cm
\epsffile{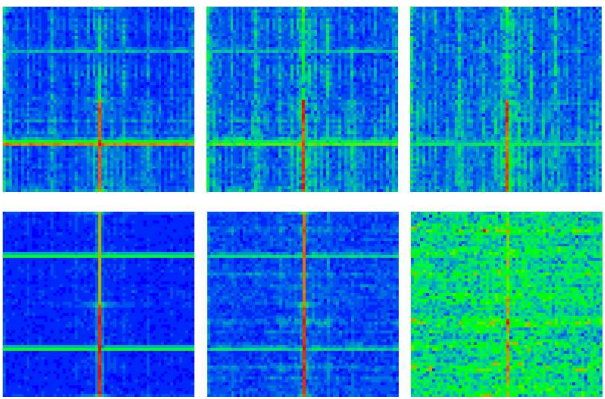}
\vglue -0.4cm
\caption{Husimi function in GA (compare with Fig.2 in \cite{grover1}),
    shown at moment $t \sim t_G$ when $w_G(t)$ has maximum , for 
    $\varepsilon= 0.002, 0.004, 0.008$ 
    (left to right respectively); $n_{tot}=12$.
    Top (bottom) row corresponds to the computation with (without)
    GYQEC at $l_g=1$. Density is proportional to color changing
    from blue/zero to red/maximum.
} 
\label{fig2}
\end{figure}

\begin{figure}
\epsfxsize=7.8cm
\epsfysize=4.8cm
\epsffile{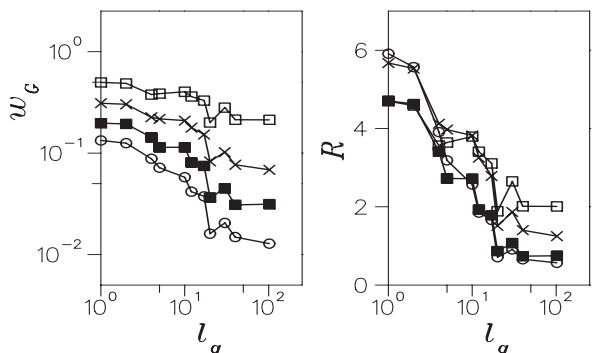}
\vglue -0.5cm
\caption{Left: probability $w_G$, averaged over 10 disorder 
realizations of static imperfections and taken at maximum,
for $\varepsilon=0.002, 0.003, 0.004, 0.005$,
computation with GYQEC at different $l_g$
(top to bottom). Right: the gain factor $R$ given by the ratio
of $w_G$ (from left) to its maximum value obtained in computations
without GYQEC (same symbols). Here $n_{tot}=12$.
} 
\label{fig3}
\end{figure}

\begin{figure}
\epsfxsize=7.3cm
\epsfysize=4.3cm
\epsffile{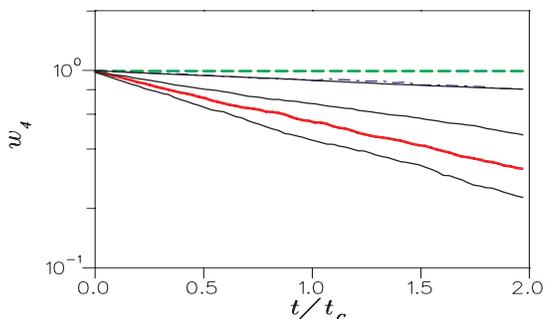}
\vglue -0.5cm
\caption{Probability  $w_4(t)$ in the four states (see Eq. (3) in \cite{grover1})
    as a function of iteration time $t$ 
    for $n_{tot}=12, \varepsilon=0.002, t_G=34.5$. 
    From top to bottom, curves show data 
    without GYQEC, with time fluctuating couplings,
    with GYQEC at $l_g=1$ (practically coincides with the previous curve)
    and at $l_g=5,10,20$.
} 
\label{fig4}
\end{figure}

GYQEC is based on a random change of numeration of qubits
after $l_g$ quantum gates. Namely, after $l_g$ gates about 
$n_{tot}/2$ swap operations are performed between 
random pairs of qubits so that the initial numeration of qubits 
is replaced by completely random one. However, in the
quantum computer code this change  is taken into account and 
the algorithm continues to run with new qubit numeration.
In a sense the method uses a freedom of numeration
of qubits in the program code and makes
gyroscopic random rotations between all possibilities. 
\newpage
These rotations suppress the effects of static imperfections
(see Fig.~1). The GA accuracy is impoved with a decrease of $l_g$ even if  at 
$l_g=1$ the GYQEC is not able to reach the case with randomly 
time fluctuating couplings between qubits \cite{note}.
A pictorial image of the accuracy improvement for the Husimi
function in shown in Fig.~2. Indeed, GYQEC gives a significant increase 
of the probability of searched state corresponding to lower horizontal line
in a phase space square in Fig.~2. 
The variation of the searched probability $w_G$ with $l_g$
is shown in Fig.~3 for various values of $\varepsilon$.
GYQEC gives a maximal improvement of accuracy 
at minimal $l_g=1$ when the effect of randomization
of static imperfections becomes maximal. For $l_g=1$ and $n_{tot}=12$
we reach the maximal accuracy gain factor $R \approx 6$
which it is not very sensitive to $\varepsilon$ in a certain range.
We expect that this $R$ value will grow with the number of qubits
$n_{tot}$ since in this case random rotations of 
computational basis will give stronger randomization of
static imperfections. We note that 
the static imperfections preserve the total probability $w_4$ in
4-states (see \cite{grover1}) until $ \varepsilon < \varepsilon_c$
while time fluctuations of couplings $a_i, b_{ij}$ and
GYQEC method  give an exponential time decay of $w_4$ 
with a rate $\Gamma \propto \varepsilon^2$ (see Fig.~4).
In spite of this decay we obtain the accuracy gain.

In summary, we discussed here a new GYQEC method
which performs random rotations in the computational basis
of quantum computation code keeping track of these rotations
in the quantum algorithm.
This method uses a generic property of numeration freedom 
in the computational basis and allows to suppress 
significantly the effects of static imperfections. 
The GYQEC is rather general and can be applied to
any quantum algorithm.

 This work was supported in part by the EC IST-FET project
EDIQIP and the NSA and ARDA under ARO contract No. DAAD19-01-1-0553.


\begin{thebibliography}{9}
\bibitem{note} For static imperfections
$w_G$ value is by factor 9 smaller compared to the case of time fluctuating couplings
(Fig.~1), that's why it's important to suppress static imperfections.
\end{thebibliography}
\end{document}